\documentclass[aps,amsfonts,amsmath,prd,preprint,nofootinbib,tightenlines]{revtex4-2}

\pdfoutput=1
\newcommand{\beq}{\begin{equation}}
\newcommand{\eeq}{\end{equation}}
\usepackage{graphicx}

\usepackage{xcolor}

\usepackage{tabularx}
\usepackage{multirow}
\usepackage{booktabs}
\usepackage{float}

\usepackage{amsmath}

\newcommand{\be}{\begin{equation}}
\newcommand{\ee}{\end{equation}}
\newcommand{\bea}{\begin{eqnarray}}
\newcommand{\eea}{\end{eqnarray}}
\newcommand{\bi}{\begin{itemize}}
\newcommand{\ei}{\end{itemize}}

\usepackage{overpic,mathtools}

\usepackage{url}
\usepackage[colorlinks]{hyperref}
\hypersetup{ 
	colorlinks=true, 
	linkcolor=black, 
	filecolor=black, 
	citecolor = blue,       
	urlcolor=blue, 
}

\begin{document}

\title{Constraining Scalar-Tensor Dark Energy with Stable Nonminimal Coupling Against DESI Data and Local Gravity Tests}
\title{Constraints on Quintessence with Stable Nonminimal Coupling\\ Using DESI Data and Solar System Tests}
\title{Constraints on Stable Nonminimally Coupled Scalar–Tensor\\ Dark Energy from DESI Data and Solar System Tests}
\title{Constraints on Stable Scalar–Tensor Dark Energy\\ from DESI Data and Solar System Tests}

\author{Husam Adam\footnote{Husam.Adam@tufts.edu}, 
Mark P. Hertzberg\footnote{mark.hertzberg@tufts.edu}, 
Daniel Jim\'enez-Aguilar\footnote{Daniel.Jimenez\_Aguilar@tufts.edu}}

\affiliation{Institute of Cosmology, Department of Physics and Astronomy, Tufts University, Medford, MA 02155, USA\\\\}

\begin{abstract}
We investigate the viability of scalar-tensor (quintessence) models of dark energy with a quartic polynomial nonminimal coupling to gravity 
and a linear scalar potential. 
The polynomial nonminimal coupling is used to ensure that the field is well stabilized in the early universe.
We perform a systematic exploration of the parameter space spanned by the quadratic and quartic nonminimal couplings, as well as the slope of the potential.
We confront the predictions of the model with the latest Dark Energy Spectroscopic Instrument (DESI) constraints on the dark energy equation of state and also with complementary bounds from local tests of gravity, including solar system constraints and limits on the time variation of the effective Newton's constant. We identify  bands in parameter space where all these constraints are satisfied, finding such bands to be very narrow. 
\end{abstract}

\maketitle

\newpage

\tableofcontents

\newpage

\section{Introduction}

The discovery that the expansion of the universe is accelerating \cite{SupernovaSearchTeam:1998fmf,SupernovaCosmologyProject:1998vns} is commonly interpreted as evidence for a dark energy component, characterized by an equation of state parameter $w_{\rm DE}=p_{\rm DE}/\rho_{\rm DE}<-1/3$ (where $p_{\rm DE}$ and $\rho_{\rm DE}$ are the dark energy pressure and density, respectively) and accounting for roughly $69\%$ of the present energy budget of the universe. Observational support for this picture comes primarily from type Ia supernovae luminosity distances and measurements of temperature anisotropies in the cosmic microwave background (CMB). Despite its phenomenological success, the fundamental nature of dark energy remains unknown.

The simplest description identifies dark energy with a cosmological constant $\Lambda$, for which the energy density is constant in time and the equation of state is fixed to $w_{\rm DE}=w_{\Lambda}=-1$. This $\Lambda$CDM framework provides an excellent fit to a wide range of cosmological observations. However, it requires an extremely small vacuum energy density $\rho_{\Lambda} \sim 10^{-47}\text{GeV}^4$ in natural units, which is in stark disagreement with theoretical expectations from quantum field theory. This enormous mismatch (often quoted as spanning more than 120 orders of magnitude when compared to the Planck scale) is known as the cosmological constant problem \cite{RevModPhys.61.1}, and will not be addressed in this work.

Recently, increasing attention has been drawn to possible deviations from the $\Lambda$CDM paradigm. In particular, results from the Dark Energy Spectroscopic Instrument (DESI) collaboration \cite{DESI:2024mwx,DESI:2025zgx} provide indications that the dark energy density may evolve with time. These analyses combine baryon acoustic oscillation (BAO) data with CMB observations \cite{Planck:2018vyg,Planck:2019nip} and several type Ia supernova compilations \cite{Rubin:2023jdq,Scolnic:2021amr,DES:2024jxu}, including Pantheon+, Union3 and DESY5, and report tensions with $\Lambda$CDM at the level of 3-4 standard deviations.

A convenient phenomenological framework to capture such deviations is the Chevallier–Polarski–Linder (CPL) parametrization \cite{Chevallier:2000qy,Linder:2002et} in which the dark energy equation of state $w$ depends linearly on scale factor as
\begin{equation}
w(a)=w_0+(1-a)w_a\,,
\label{eq:CPL}
\end{equation}
where $a$ denotes the scale factor (normalized to 1 today) and $w_0$ and $w_a$ represent the current value of the equation of state parameter and its first derivative, respectively. The cosmological constant $w=-1$ corresponds to the limit $w_0=-1$ and $w_a=0$. 

Interestingly, the central values reported by DESI, approximately given by $w_0\approx -0.7$ and $w_a\approx-1$ for the combined DESI BAO + CMB + supernovae data (see Fig. \ref{fig:contours}), suggest an evolution of the form $w_{\rm DE}(a)\approx -1.7 + a$. If extrapolated to earlier times, this would imply $w_{\rm DE}<-1$ for $a\lesssim 0.7$ (or redshift $z\gtrsim 0.43$), indicating a violation of the null energy condition. Such behavior cannot be realized within minimally coupled scalar field models with canonical kinetic terms. For these models, $\rho_{\rm DE}+p_{\rm DE}\geq 0$ necessarily holds, implying that $w_{\rm DE}\geq-1$. Since $z\gtrsim 0.43$ overlaps with the range of redshifts probed by DESI ($0.295\leq z \leq 2.33$), these models are expected to provide, at best, a modest fit to the data. A non-exhaustive list of works in which canonical quintessence models are studied in light of the recent DESI data is \cite{Bhattacharya:2024hep,Berghaus:2024kra,Ramadan:2024kmn,Wolf:2024eph,Andriot:2024sif,Bhattacharya:2024kxp,Akrami:2025zlb,Dinda:2025iaq,deSouza:2025rhv,Gialamas:2025pwv,Bayat:2025xfr,Cline:2025sbt,Adam:2025kve,Shlivko:2025krk}. 
This observation motivates the exploration of more general frameworks for dynamical dark energy, including interacting dark sector models \cite{Wang:2024vmw,Li:2025owk,Chakraborty:2025syu,vanderWesthuizen:2025iam,Guedezounme:2025wav,Samanta:2025oqz,Khoury:2025txd,Pan:2025qwy,LaPenna:2026avs,Antusch:2026ldp,Gomez-Valent:2026ept} or scalar-tensor theories with nonminimal couplings \cite{Ye:2024ywg,Wolf:2024stt,Ferrari:2025egk,Wolf:2025jed,Wang:2025znm,Pan:2025psn,Adam:2025kve,Wang:2025znm,SanchezLopez:2025uzw,Li:2026aqz} (see also \cite{Wang:2026wrk}). In this work, we will focus on the latter possibility.

In Ref.~\cite{Adam:2025kve}, we exemplified how a theory consisting of a scalar field nonminimally coupled to gravity could be compatible with the results reported by DESI. For careful choices of the scalar potential, the initial conditions and the value of the nonminimal coupling to the Ricci scalar, we found cases in which the equation of state parameter provided a good fit to the data. Furthermore, the tight bounds on the fifth force and the time variation of Newton’s gravitational constant $G$ placed by different tests of gravity were avoided. Nevertheless, one of the main drawbacks of the model is the fact that the dark energy turned out to be the dominant form of energy at the time of recombination and earlier, thus spoiling the standard dynamics of the early universe. In the present work, we address this issue by extending the theory in such a way that the nonminimal coupling term in the Lagrangian takes the form $\xi f(\varphi) R$ for some polynomial function $f$ of the scalar field $\varphi$, instead of the usual $\xi\varphi^{2}R$. As we shall see, apart from keeping the evolution of the early universe under control, this approach also allows one to define a natural initial condition for the field, thereby alleviating to some extent the fine tuning needed in the previous model.

The manuscript is organized as follows. In section \ref{sec:model} we describe the model under consideration, consisting of a real scalar field with polynomial nonminimal coupling to gravity. In section \ref{sec:constraints} we present the different bounds imposed on the evolution of the field by local tests of gravity. In section \ref{sec:ic}, we specify the initial conditions used in the numerical integration of the Friedmann and Klein-Gordon equations. We also analyze in more detail the shortcomings of the model considered in \cite{Adam:2025kve}, showing how they are resolved by the introduction of the polynomial nonminimal coupling. 
In section \ref{sec:results} we present our numerical results.
Finally, in section \ref{sec:conclusions} we discuss and conclude.

\section{Model}
\label{sec:model}
The model we are considering in this study is given by the action (signature $-+++$ and units $\hbar=c=1)$
\begin{equation}
S=\int d^{4}x\sqrt{-g}\left[\frac{M_{p}^{2}}{2}R-\frac{1}{2}g^{\mu\nu}\partial_{\mu}\varphi\partial_{\nu}\varphi-V(\varphi)-\frac{1}{2}\xi f(\varphi)R+\mathcal{L}_{\rm m}\right]\,,
\label{eq:action}
\end{equation}
where $M_p=1/\sqrt{8\pi G}$, $\mathcal{L}_{\rm m}$ denotes the Lagrangian density of the fields that describe matter, $\xi$ is the nonminimal coupling parameter, and
\begin{equation}
f(\varphi)=\varphi^{2}-\frac{\gamma}{2}\varphi^{4}\,,
\label{eq:f}
\end{equation}
with $\gamma$ a positive constant with dimensions of $E^{-2}$ (henceforth, $E$ refers to energy). This choice of coupling function will be explained in section \ref{sec:ic}. We will consider a potential energy density of the form
\begin{equation}
V(\varphi)=V_{0}+C\varphi\,,
\label{eq:V}
\end{equation}
where $V_{0}$ and $C$ are constants with dimensions of $E^{4}$ and $E^{3}$, respectively. The absence of a linear term in $f$ is possible without loss of generality by shifting $\varphi$ to ensure this. Generically, this requires the above constant $V_0$ in $V$ to be included.
Now, the assumption of a linear potential is normally justified when the field experiences a sufficiently small excursion during its evolution. 
However, we will at times consider somewhat large field excursions, so this linear choice is not guaranteed to be justified. So we take it as merely a representative example. More general potentials could be considered elsewhere (for example, in Ref.~\cite{Adam:2025kve}, we considered a range of potentials with the quadratic nonminimal coupling).
Altogether, taking into account the nonminimal coupling to gravity, the scalar field evolves under the influence of an effective potential
\begin{equation}
V_{\rm eff}(\varphi)=V(\varphi)+\frac{1}{2}\xi f(\varphi)R\,.
\label{eq:V eff}
\end{equation}

Assuming that the field is homogeneous, the action (\ref{eq:action}) leads to the following equation of motion
\begin{equation}
\ddot{\varphi}+3H\dot{\varphi}+V'(\varphi)+\frac{1}{2}\xi f'(\varphi)R=0\,,
\label{eq:eom}
\end{equation}
as well as the Friedmann equation for the evolution of the scale factor\footnote{We are considering a spatially flat Friedmann-Lemaître-Robertson-Walker universe.}:
\begin{equation}
H^{2}=\left(\frac{\dot{a}}{a}\right)^{2}=\frac{1}{3M_{p}^{2}}(\rho_{m}+\rho_{\varphi})\,.
\label{eq:Friedmann}
\end{equation}
Here, $\rho_{m}$ and $\rho_{\varphi}$ are the energy densities of matter and the scalar field. The latter corresponds to the time-evolving dark energy density.

The energy-momentum tensor (computed as $T_{\mu\nu}=-(2/\sqrt{-g})\,\delta(\sqrt{-g}\mathcal{L}_{\varphi})/\delta g^{\mu\nu}$, where $\mathcal{L}_{\varphi}$ is the Lagrangian density of the field) reads
\begin{equation}
T_{\mu\nu}=\partial_{\mu}\varphi\partial_{\nu}\varphi-g_{\mu\nu}\left[\frac{1}{2}g^{\alpha\beta}\partial_{\alpha}\varphi\partial_{\beta}\varphi+V(\varphi)\right]+\xi\left(G_{\mu\nu}+g_{\mu\nu}\Box-\nabla_{\mu}\nabla_{\nu}\right)f(\varphi)\,,
\label{eq:Tmunu}
\end{equation}
where $G_{\mu\nu}$ is the Einstein tensor. From this expression, one can compute the energy density $\rho_\varphi$ and pressure $p_\varphi$ of the field:
\begin{equation}
\rho_{\varphi}=\frac{1}{2}\dot{\varphi}^{2}+V(\varphi)+3H\xi\left[Hf(\varphi)+f'(\varphi)\dot{\varphi}\right]\,,
\label{eq:energy density phi2}
\end{equation}
\begin{equation}
p_{\varphi}=\frac{1}{2}\dot{\varphi}^{2}-V(\varphi)
+\xi\left\{-f''(\varphi)\dot{\varphi}^{2}+f'(\varphi) V'(\varphi)+\left(H^{2}-\frac{R}{3}\right)f(\varphi)+\frac{\xi R}{2}\left[f'(\varphi)\right]^{2}+Hf'(\varphi)\dot{\varphi}\right\}\,.
\label{eq:pressure phi}
\end{equation}
(See ahead to Section \ref{PhysObs} for closely related quantities, $\bar\rho_\varphi$ and $\bar{p}_\varphi$, which are the physically measurable ones.)

\section{Constraints}
\label{sec:constraints}
The nonminimal coupling of the scalar field to gravity has specific consequences on the evolution of the universe on both small and large scales. The non-observation of these effects places tight bounds on the dynamics of the scalar field.

\subsection{Gravitational constraint}

The most obvious consequence of the nonminimal coupling is, perhaps, the fact that it leads to an effective gravitational constant $G_{\rm eff}$ which is time-dependent. From the action (\ref{eq:action}) it is easily seen that
\begin{equation}
G_{\rm eff}=\frac{G}{1-\frac{\xi f(\varphi)}{M_{p}^{2}}}\,.
\label{eq:G effective}
\end{equation}
A time-varying Newton's constant modifies the orbits of planets and satellites over time with respect to the prediction of general relativity, and it also alters the universe's expansion rate at different redshifts, potentially affecting the physical processes taking place in the early universe, such as nucleosynthesis and last scattering.

Bounds on the time variation of the gravitational coupling come from lunar and planetary ranging experiments, consisting of high-precision measurements of the orbits of various bodies of the solar system, as well as from the analysis of the timing data of pulsars (see section 6 in \cite{Uzan:2024ded} and references therein). These experiments impose
\begin{equation}
\Bigg|\frac{\dot{G}_{\rm eff}}{G_{\rm eff}}\Bigg|_{0}\lesssim10^{-12}\text{\, yr}^{-1}\,,
\label{eq:Gdot over G bound}
\end{equation}
where the subscript 0 indicates evaluation at the present cosmic time $t_{0}$. Using (\ref{eq:G effective}), this translates into
\begin{equation}
\Bigg|\frac{\xi f'(\varphi)\dot{\varphi}}{M_{p}^{2}-\xi f(\varphi)}\Bigg|_{0}\lesssim10^{-12}\text{\, yr}^{-1}\,.
\label{eq:Gdot over G bound 2}
\end{equation}
Here we are taking a conservative bound. We note that there are claims of tighter bounds towards $10^{-13}\text{\, yr}^{-1}$ and $10^{-14}\text{\, yr}^{-1}$. We will call this the ``gravitational constraint".


\subsection{Solar system constraint}
Consider small perturbations of the scalar field around its current value: $\varphi=\varphi_{0}+\delta\varphi$. Expanding the nonminimal coupling term in the action to first order in $\delta\varphi$, one identifies a term proportional to $\delta\varphi R$ in the perturbed Lagrangian. Since 
\begin{equation}
R=\frac{\rho_{\rm m}+\rho_{\varphi}-3p_{\varphi}}{M_{p}^{2}}\,,
\end{equation}
the aforementioned term reveals the presence of a fifth force between non-relativistic matter mediated by the $\delta\varphi$ field. The coupling between this scalar and the trace of the energy-momentum tensor of the matter fields is given by $y=\xi f'(\varphi)/2M_{p}^{2}$, and it is usually constrained via the post-parametrized Newtonian parameter $\gamma_{_{\rm PPN}}$, which is introduced in the weak-field metric $ds^2=-(1+2\phi_N)dt^2+(1+2\gamma_{_{\rm PPN}}\phi_N)|d{\bf x}|^2$. Here, $\phi_{N}$ is the Newtonian gravitational potential.
This yields (assuming the physical Newton's constant $G_N=G_{\rm eff,0}$ is close to $G=1/(8\pi M_p^2)$; see 
Eq. (\ref{eq:G effective}) for the mismatch)
\begin{equation}
    \gamma_{_{\rm PPN}}={1-2y^2 M_p^2\over 1+2y^2M_p^2}\,.
\end{equation}
The $\gamma_{_{\rm PPN}}$ parameter is strongly bounded by measurements of the Shapiro time delay of radio signals from the Cassini spacecraft \cite{Bertotti:2003rm} (see also \cite{Karam:2026sqg}): $|\gamma_{_{\rm PPN}}-1|<2.3\times 10^{-5}$. With the above information, this leads to
\begin{equation}
\frac{|\xi f'(\varphi_{0})|}{4.85\times 10^{-3}M_{p}}<1\,.
\label{eq:solar system constraint 1}
\end{equation}
We will call this the ``solar system constraint". (Note that both this and the previous one involve observations in the solar system, and so they are both types of solar system tests.)

\section{Initial conditions and early-universe dynamics}
\label{sec:ic}
In the following section, we will present the results obtained for the evolution of the scalar field in this theory and their compatibility with the latest DESI data\footnote{We choose to work in the Jordan frame. See section IV C in \cite{Adam:2025kve} and references therein for a brief discussion of the suitability of this choice to test theoretical predictions against observational data.} \cite{DESI:2025zgx}. First of all, in order to solve Eqs. (\ref{eq:eom}) and (\ref{eq:Friedmann}) numerically, we need to make all variables dimensionless. 

The energy density will be computed in units of the constant $V_{0}$ appearing in (\ref{eq:V}). Without loss of generality, we set it to $V_{0}=m^{2}M_{p}^{2}$, where $m$ is some mass (or energy) scale. Therefore, the dimensionless potential energy density, $\tilde{V}(\varphi)\equiv V(\varphi)/V_{0}$, is given by
\begin{equation}
\tilde{V}(\varphi)=1+\beta\tilde{\varphi}\,,
\end{equation}
where we have defined the dimensionless constant $\beta=C/(m^2M_{p})$. Furthermore, if we rescale the field as $\tilde{\varphi}=\varphi/M_{p}$, the equation of motion will be free of parameters provided that the space and time coordinates, as well as the constant $\gamma$, are redefined as $\tilde{x}^{\mu}\equiv mx^{\mu}$ and $\tilde{\gamma}\equiv M_{p}^{2}\gamma$. Then, the dimensionless Ricci scalar and Hubble rate are $\tilde{R}=R/m^{2}$ and $\tilde{H}=H/m$, and the coupling function $\tilde{f}=f/M_{p}^2$ takes the same form (\ref{eq:f}) with tildes on the right-hand side.

In the very early universe, the effective potential (\ref{eq:V eff}) is expected to be dominated by the term proportional to the Ricci scalar. Taking this into account, and given our choice of coupling function $f(\varphi)$ in Eq. (\ref{eq:f}), we assume that the field starts out at 
\begin{equation}
\varphi_{i}=\frac{1}{\sqrt{\gamma}}\,,
\label{eq:initial phi}
\end{equation}
which is the value of $\varphi$ that minimizes $f(\varphi)$ and thus the initial effective potential. The initial field velocity is set to zero.

We solve simultaneously the dimensionless version of Eqs. (\ref{eq:eom}) and (\ref{eq:Friedmann}) with these initial conditions for the field and an initial matter density of 
\begin{equation}
\rho_{\text{m},i}=10^{9}V_{0}
\end{equation}
so that the initial dark energy density is subdominant.

\subsection{Quadratic coupling $f(\varphi)=\varphi^2$}

As briefly mentioned in the introduction, the particular example presented in \cite{Adam:2025kve}, where $f(\varphi)=\varphi^{2}$, was problematic. In that case, for the choice $\beta=1$, $\xi=-0.506$, $\varphi_{i}=0.08M_{p}$ and $\dot{\varphi}_{i}=0$, as well as $\rho_{\text{m},i}=10^{6}\rho_{\varphi,i}$, we showed how the dark energy equation of state predicted by the model provided a good fit to the DESI data. Additionally, the gravitational and solar system constraints were satisfied due to an ``accidental'' smallness of the current value of $\varphi$. However, at redshifts higher than the one corresponding to the initial conditions, the dark energy density quickly overcame the matter density, spoiling the standard dynamics of the early universe. This unstable behavior of dark energy at early times can be easily understood by inspecting the equation of motion for the field. At the time corresponding to our initial state, matter is the dominant form of energy. Therefore, the Hubble rate and the Ricci scalar are approximately given by $H\approx 2t^{-1}/3$ and $R\approx 4t^{2}/3$. Substituting into (\ref{eq:eom}), and taking into account that the term proportional to the Ricci scalar is the dominant contribution to the effective potential, one finds
\begin{equation}
3t^{2}\ddot{\varphi}+6t\dot{\varphi}+4\xi\varphi=0\,,
\end{equation}
which has the general solution
\begin{equation}
\varphi(t)=C_{1}t^{\epsilon_{-}}+C_{2}t^{\epsilon_{+}}\,,
\end{equation}
with
\begin{equation}
\epsilon_{\pm}=\frac{-3\pm\sqrt{9-48\xi}}{6}\,.
\end{equation}

\begin{figure}[t!]
    \centering
    \includegraphics[scale=0.8]{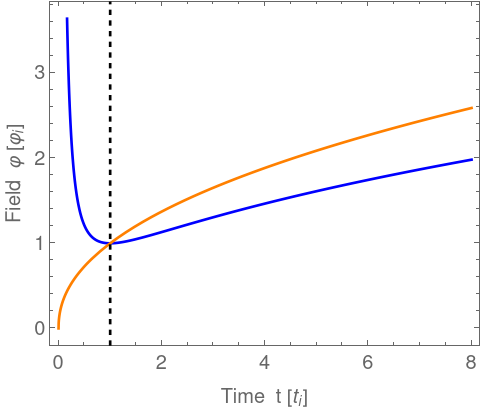}
    \caption{Behavior of the scalar field in the matter-dominated universe (see Eq. (\ref{eq:decaying and growing modes})). The blue curve corresponds to an initially static configuration ($v_{i}=0$), while the orange one corresponds to the fine-tuned initial velocity $v_{i}=\epsilon_{+}$. The dashed line corresponds to the initial time $t_{i}$.}
    \label{fig:unstable phi}
\end{figure}

If we introduce the dependence of the constants $C_{1,2}$ on the initial conditions and rescale the field and cosmic time by their initial values, $\phi\equiv\varphi/\varphi_{i}$ and $\tau\equiv t/t_{i}$, we get
\begin{equation}
\phi(\tau)=\frac{\epsilon_{+}-v_{i}}{\epsilon_{+}-\epsilon_{-}}\tau^{\epsilon_{-}}-\frac{\epsilon_{-}-v_{i}}{\epsilon_{+}-\epsilon_{-}}\tau^{\epsilon_{+}}\,,
\label{eq:decaying and growing modes}
\end{equation}
where $v_{i}$ is $d\phi/d\tau$ evaluated at the initial time. In this particular example, $v_{i}=0$. Also, since $\xi=-0.506$, we have $\epsilon_{-}\approx-1.46$ and $\epsilon_{+}\approx 0.46$. Therefore, the field is the sum of a decaying mode and a growing mode. Since the former dominates at early times, the field grows at higher redshifts and the dark energy density blows up. This is illustrated in Fig. \ref{fig:unstable phi}. This behavior can be traced to the negative sign of the nonminimal coupling constant $\xi$, which effectively makes the field tachyonic. Note, however, that the decaying mode can be eliminated by the choice of initial velocity $v_{i}=\epsilon_{+}$. With this fine-tuning, the dark energy density would remain subdominant at higher redshifts.

\subsection{Quartic coupling $f(\varphi)=\varphi^2-\gamma\,\varphi^4/2$}

The introduction of the quartic term in the function $f(\varphi)$, controlled by the parameter $\gamma$, allows us to avoid this instability. If the scalar field starts out at the minimum of $f(\varphi)$, which very approximately corresponds to the minimum of the effective potential, it will remain there at earlier times; it is {\em stable}.

\begin{figure}[h!]
\centering
\includegraphics[scale=0.67]{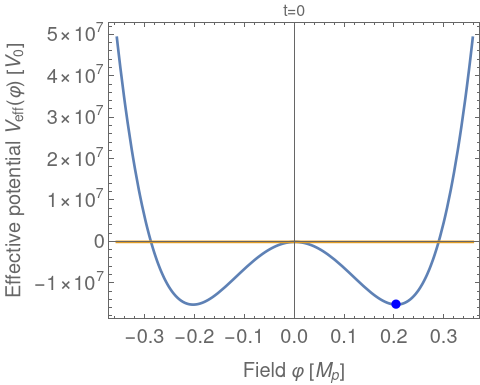}
\includegraphics[scale=0.61]{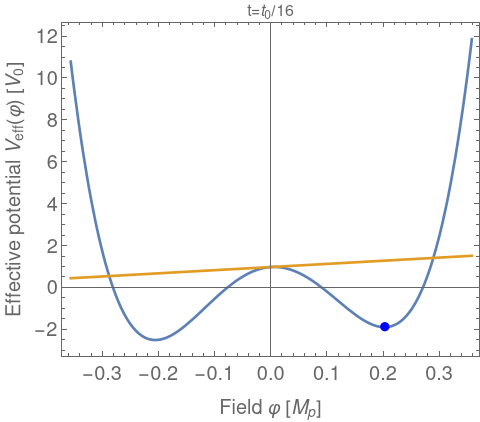}
\includegraphics[scale=0.61]{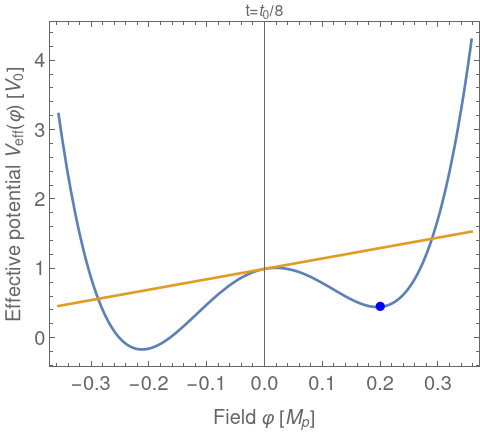}
\includegraphics[scale=0.61]{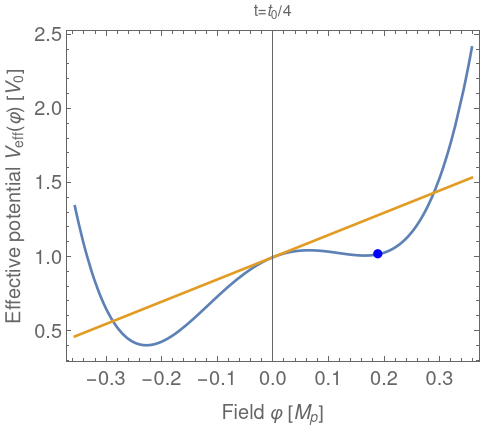}
\includegraphics[scale=0.61]{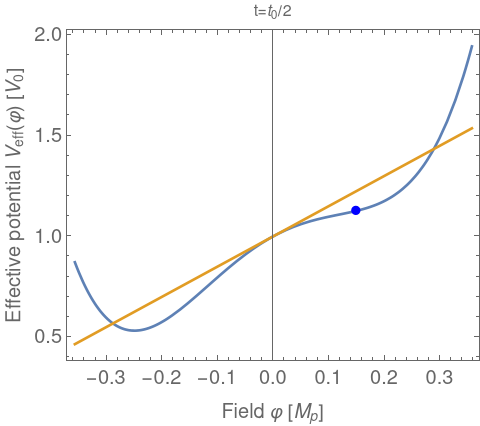}
\includegraphics[scale=0.61]{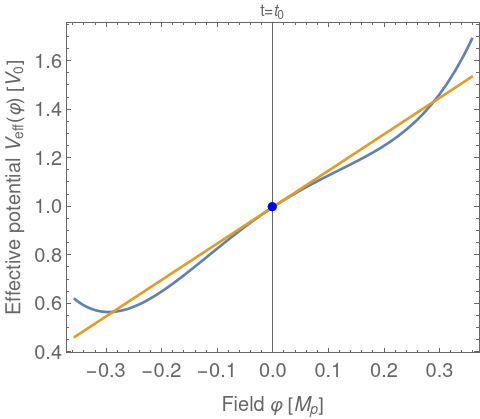}
\caption{Evolution of the scalar field (blue dot) on the effective potential (blue curve) for $\beta=1.5$, $\xi=-1.5$ and $\tilde\gamma=24$. The orange line represents the potential $V(\varphi)$. (Note that while the potential $V$ is fixed, it appears to change slope as we zoom in on the relevant scales.)}
\label{fig:effective potential snapshots}
\end{figure}

In Fig. \ref{fig:effective potential snapshots}, we show some snapshots of the evolution of the scalar field on the effective potential (\ref{eq:V eff}). At early times, the term proportional to the Ricci scalar dominates the dynamics. Consequently, the effective potential near the
initial condition (\ref{eq:initial phi}) is a stable local minimum, ensuring the stability of the scalar field at high redshifts. At later times, as the universe expands and $R$ decreases, that term decreases in absolute value and the effective potential $V_{\rm eff}$ approaches the linear potential $V$. 

\section{Results}
\label{sec:results}
\begin{figure}[t!]
    \centering
    \includegraphics[scale=0.3]{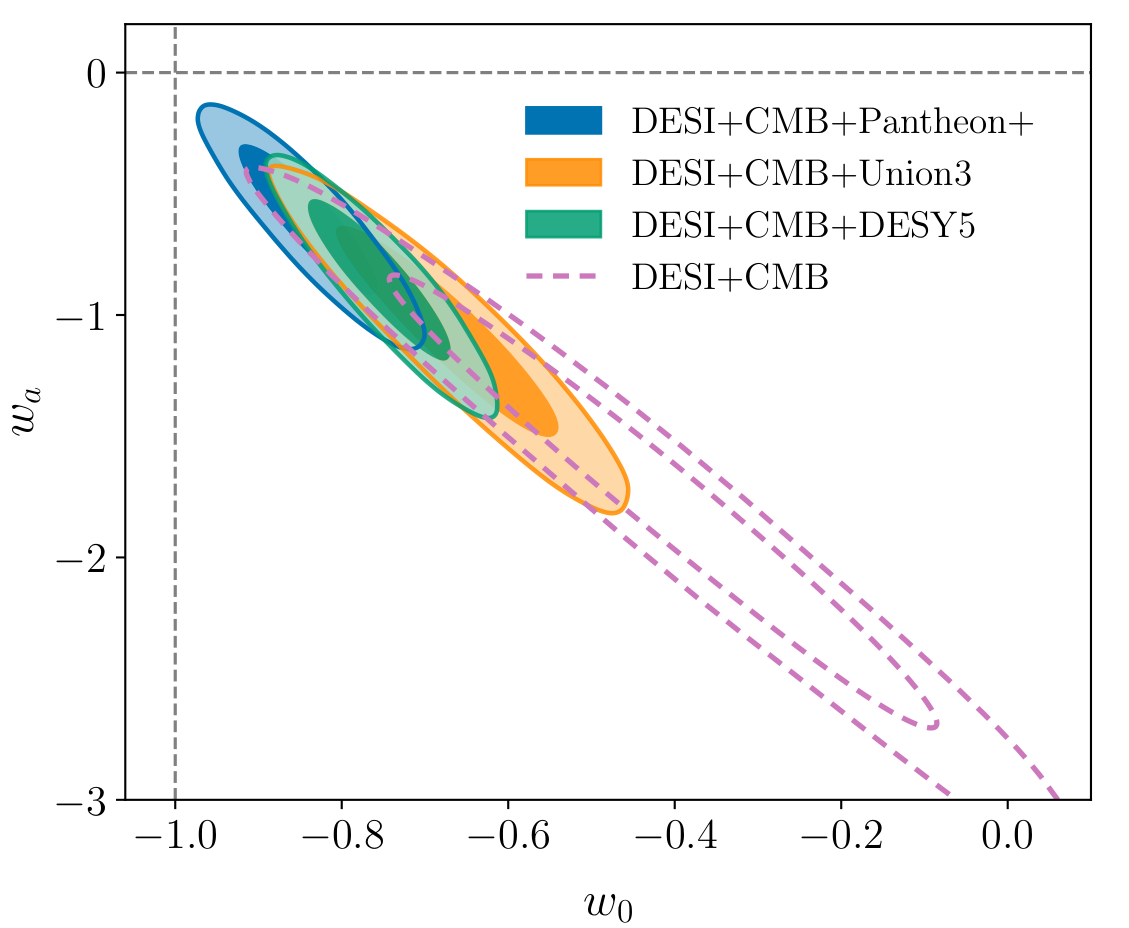}
    \caption{$2\sigma$ DESI 2025 contours for different combinations of BAO, CMB and type IA supernovae datasets. This image was taken directly from Ref.~\cite{DESI:2025zgx}.}
    \label{fig:contours}
\end{figure}

Once we have fixed the initial conditions, only $\xi$, $\tilde\gamma$ and $\beta$ remain free. We varied these three parameters over the following ranges and step sizes:
\begin{equation}
\xi\in[-3,1]\,\,\,,\,\,\,\Delta\xi=0.1\,\,\,;
\label{eq:xi range}
\end{equation}
\begin{equation}
\tilde{\gamma}\in[0.1,30]\,\,\,,\,\,\,\Delta\tilde{\gamma}=0.1\,\,\,;
\label{eq:gamma range}
\end{equation}
\begin{equation}
\beta\in[0.25,2]\,\,\,,\,\,\,\Delta\beta=0.25\,\,\,.
\label{eq:beta range}
\end{equation}
This scan consists of a total of 98,400 different combinations of parameters, or 12,300 different simulations for each specific value of $\beta$. In each of these realizations, we computed the dark energy equation of state as a function of time and found the best-fit $w_{0}$ and $w_{a}$ parameters using the CPL parametrization (\ref{eq:CPL}) over the DESI BAO range of redshifts $0.295\leq z\leq 2.33$, and then checked whether the obtained pair of values ($w_{0}$,$w_{a}$) fell within the $2\sigma$ DESI contours shown in Fig. \ref{fig:contours}. In order to do this, we first extracted the coordinates of the outer ($2\sigma$) contours and fitted them to ellipses of the form
\begin{equation}
F_{i}(w_{0},w_{a})=I_{i}+J_{i}w_{0}+K_{i}w_{a}+L_{i}w_{0}w_{a}+M_{i}w_{0}^{2}+N_{i}w_{a}^{2}\,,
\label{eq:ellipse fit}
\end{equation}
where $I_{i},J_{i},K_{i},L_{i},M_{i},N_{i}$ are real numbers and the subindex $i=1,2,3,4$ labels the particular combination of datasets, with $F_i=0$ being the ellipse. Then, we simply checked if the particular point ($w_{0}$,$w_{a}$) obtained in the simulation was inside any of the ellipses, that is, if
\begin{equation}
F_{i}(w_{0},w_{a})<0\,\,\,\,\text{for }\,i=1,2,3\text{ or }4.
\label{eq:ellipse constraint}
\end{equation}
In the following, we will refer to this condition as the ``ellipse constraint''.

Note that we are considering the combined DESI BAO + CMB + supernovae contours even if the CPL parameters are found by fitting the dark energy equation of state in the range of redshifts that correspond exclusively to the BAO survey ($0.295\leq z\leq 2.33$). However, since dark energy contributes weakly to the angular diameter distance to the last scattering surface at high redshifts, the likelihood associated with the CMB survey does not constrain the redshift dependence of the equation of state at such early times (see \cite{Wolf:2024eph}). Therefore, our approach should be accurate enough.

\subsection{Physical Observables}
\label{PhysObs}

In order to compare the results of our model to the DESI data, we need to define quantities with respect to Newton's constant evaluated today $G_N=G_{\rm eff,0}\approx 6.67 \times 10^{-11}\,\rm{m}^3/\rm{kg}/\rm{s}^2$
(see Eq. (\ref{eq:G effective}) with $\varphi=\varphi_0$).
To do so, we rewrite the Friedmann equation as
\begin{equation}
H^{2}=\frac{8\pi G_{\rm eff,0}}{3}(\rho_{\rm m}+\bar{\rho}_{\varphi})\,.
\end{equation}
From this we see that the modified dark energy density $\bar{\rho}_{\varphi}$ is related to the earlier defined densities from Eq. (\ref{eq:Friedmann}) by
\begin{equation}
\bar{\rho}_{\varphi}=\left(\frac{G}{G_{\rm eff,0}}-1\right)\rho_{\rm m}+\frac{G}{G_{\rm eff,0}}\rho_{\varphi}\,.
\label{eq:modified rho}
\end{equation}
Applying the same logic to the acceleration equation,
\begin{equation}
\frac{\ddot{a}}{a}=-\frac{4\pi G}{3}(\rho_{\rm m}+\rho_{\varphi}+3p_{\varphi})=-\frac{4\pi G_{\rm eff,0}}{3}(\rho_{\rm m}+\bar{\rho}_{\varphi}+3\bar{p}_{\varphi})\,,
\end{equation}
we get the modified dark energy pressure
\begin{equation}
\bar{p}_{\varphi}=\frac{G}{G_{\rm eff,0}}p_{\varphi}\,.
\label{eq:modified p}
\end{equation}
Therefore, the equation of state is
\begin{equation}
w=\frac{\bar{p}_{\varphi}}{\bar{\rho}_{\varphi}}=\frac{p_{\varphi}}{\rho_{\varphi}+\left(1-\frac{G_{\rm eff,0}}{G}\right)\rho_{\rm m}}\,.
\label{eq:modified w}
\end{equation}
By solving for this numerically and then obtaining a linear fit to this, we can extrapolate the CPL parameters of Eq. (\ref{eq:CPL}).

One can also define the modified density fractions of matter and dark energy as
\begin{equation}
\bar{\Omega}_{\rm m}=\frac{\rho_{\rm m}}{\rho_{\rm m}+\bar{\rho}_{\varphi}}=\frac{G_{\rm eff,0}}{G}\Omega_{\rm m}
\label{eq:modified Omega matter}
\end{equation}
and
\begin{equation}
\bar{\Omega}_{\varphi}=\frac{\bar{\rho}_{\varphi}}{\rho_{\rm m}+\bar{\rho}_{\varphi}}=1-\frac{G_{\rm eff,0}}{G}\Omega_{\rm m}\,,
\label{eq:modified Omega phi}
\end{equation}
where $\Omega_{\rm m}=\rho_{\rm m}/(\rho_{\rm m}+\rho_{\varphi})$ is the un-modified matter density fraction.
The cosmic time corresponding to the present day, $t_{0}$, is found as the time at which $\bar{\Omega}_{\rm m}=0.31$ (or $\bar{\Omega}_{\varphi}=0.69$).
However, since this typically happens twice in the time ranges we numerically explored, we define $t_{0}$ as the first time that the condition is satisfied.\footnote{As a numerical trick, this cosmic time can be found by numerically solving the equation
\begin{equation}
\frac{\Omega_{\rm}(t)}{1-\frac{\xi f\left[\phi(t)\right]}{M_{p}^{2}}}+\frac{1}{2}\left[\frac{\dot{\Omega}_{\rm m}(t)}{|\dot{\Omega}_{\rm m}(t)|}+1\right]=0.31\,.
\end{equation}
Since matter dominates initially, its density fraction has negative time-derivative when it hits 0.31 for the first time. In this case, the second term on the left-hand side vanishes and the solution $t_{0}$ satisfies the desired condition $\bar{\Omega}_{\rm m}(t_{0})=0.31$, while on the second pass the matter fraction has positive time-derivative and the second term is 1, and there is no solution of this equation.}

\begin{figure}[h!]
\centering
\includegraphics[scale=0.57]{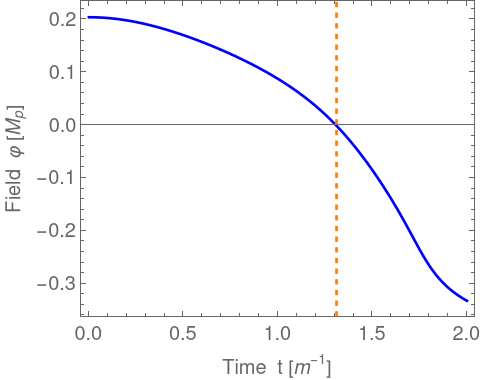}
\hspace{1cm}
\includegraphics[scale=0.55]{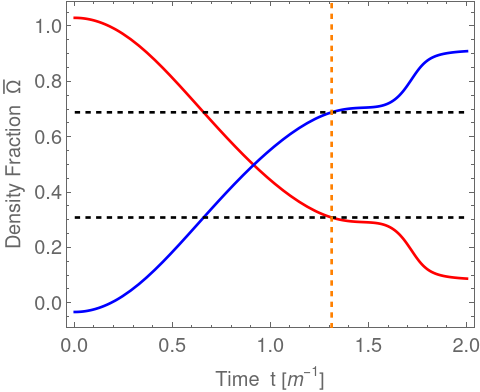}\\
\vspace{1cm}
\includegraphics[scale=0.57]{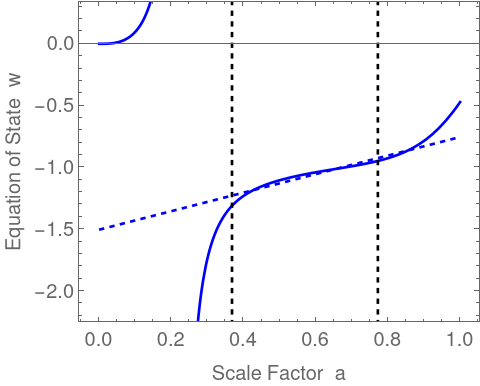}
\hspace{1cm}
\includegraphics[scale=0.56]{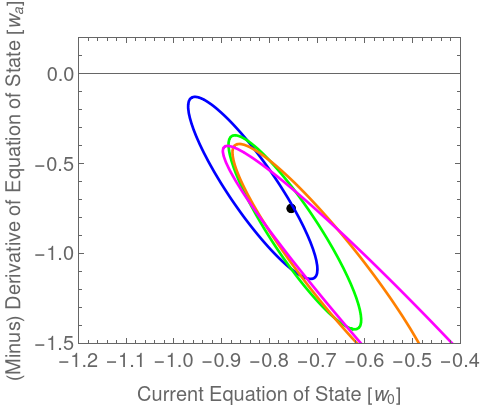}
\caption{Top: scalar field and density fractions of dark energy (blue) and matter (red), as given by Eqs. (\ref{eq:modified Omega phi}) and (\ref{eq:modified Omega matter}) as functions of cosmic time. The orange dashed line indicates the current time $t_{0}$, and the horizontal black lines mark the current values of the density fractions (0.69 for dark energy and 0.31 for matter). Bottom: dark energy equation of state parameter (as given by (\ref{eq:modified w})) as a function of the scale factor. The blue dashed line in this plot corresponds to the CPL fit over the relevant range of redshifts, which in turn are represented by the vertical black lines. The corresponding ($w_{0}$,$w_{a}$) point is shown as a black dot in the DESI contour plot.}
\label{fig:more plots}
\end{figure}

The value of the scalar field as a function of cosmic time, as well as the equation of state parameter (\ref{eq:modified w}) and the density fractions of matter and dark energy (\ref{eq:modified Omega matter}) and (\ref{eq:modified Omega phi}), are shown in Fig. \ref{fig:more plots} for a particular simulation with $\beta=1.5$, $\xi=-1.5$ and $\tilde\gamma=24$. In this case, note that the equation of state exhibits phantom crossing and is almost linear in the DESI range of redshifts. The corresponding CPL parameters $w_{0}$ and $w_{a}$ lie well within the $2\sigma$ contours. Furthermore, as will become evident in the following subsection, the small present-day value of the scalar field favors compliance with the gravitational and solar system constraints. However, as we will note, this is not the generic situation.   

\subsection{Constraints}

Apart from locating the dark energy equation of state parameters in DESI's $w_{0},w_{a}$ plane, we checked whether the gravitational and solar system constraints (\ref{eq:Gdot over G bound 2}) and (\ref{eq:solar system constraint 1}) were satisfied or not. 
We did this over 98,400 runs of parameters.

In order to perform this analysis, it is useful to define dimensionless versions of these conditions.
Let us start with the constraint on the effective gravitational coupling. Using the dimensionless variables defined above and dividing both sides of (\ref{eq:Gdot over G bound 2}) by $m=H_{0}/\tilde{H}_{0}$ (where $H_{0}\approx 6.9\times 10^{-11}\,\text{yr}^{-1}$ is the observed value of the Hubble rate today \cite{Planck:2018vyg} and $\tilde{H}_{0}$ is its dimensionless value directly extracted from the simulation), we get
\begin{equation}
\Bigg|\frac{\xi \tilde{f}'(\tilde{\varphi})\dot{\tilde{\varphi}}}{1-\xi \tilde{f}(\tilde{\varphi})}\Bigg|_{0}\lesssim 0.014\tilde{H}_{0}\,,
\label{eq:Gdot over G bound 3}
\end{equation}
where dots and primes denote now derivatives with respect to the dimensionless cosmic time $\tilde{t}$ and field $\tilde{\varphi}$, respectively. If we define the function
\begin{equation}
g(\tilde{t})\equiv\frac{\tilde{H}^{-1}(\tilde{t})}{0.014}\,\frac{\xi \tilde{f}'(\tilde{\varphi})\dot{\tilde{\varphi}}}{1-\xi \tilde{f}(\tilde{\varphi})}\,,
\label{eq:g of t}
\end{equation}
the gravitational constraint finally reads
\begin{equation}
-1<g(\tilde{t}_{0})<1\,.
\label{eq:gravitational constraint}
\end{equation}
Similarly, by defining the function
\begin{equation}
s(\tilde{t})\equiv\frac{\xi\tilde{f}'(\tilde{\varphi})}{4.85\times 10^{-3}}\,,
\label{eq:s of t}
\end{equation}
the solar system constraint (\ref{eq:solar system constraint 1}) is
\begin{equation}
-1<s(\tilde{t}_{0})<1\,.
\label{eq:solar system constraint}
\end{equation}

\begin{figure}[h!]
\centering
\includegraphics[scale=0.88]{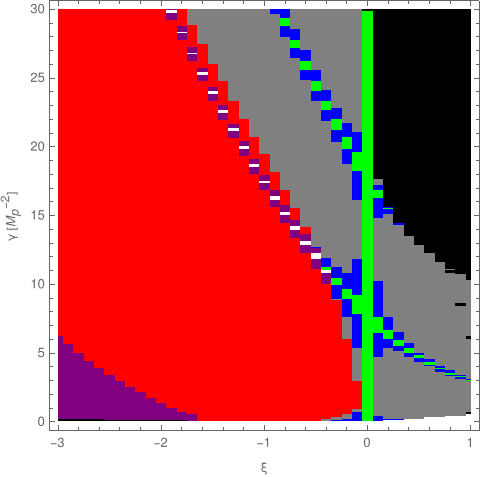}\\
\vspace{0.5cm}
\hspace{1cm}
\includegraphics[scale=0.4]{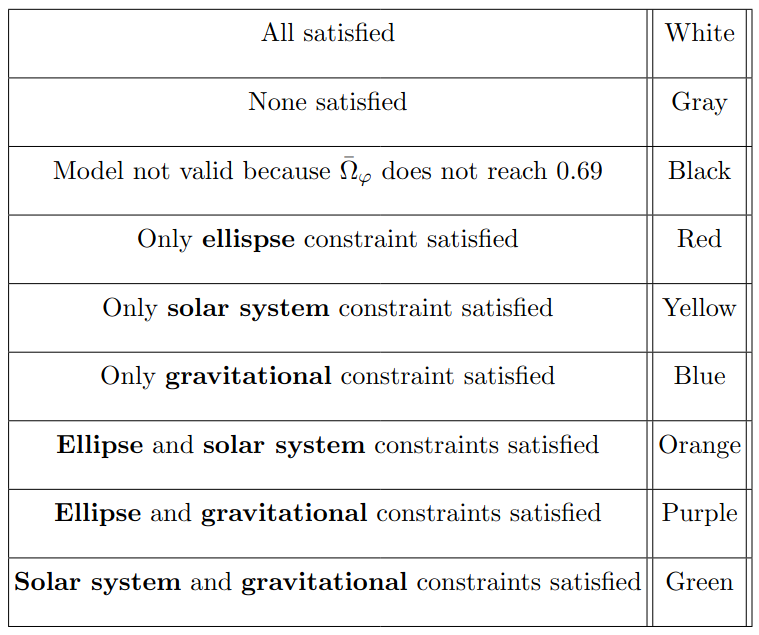}
\caption{Constraints that are satisfied 
depending on $\xi$ and $\tilde\gamma$, for $\beta=1.5$.
The ellipse, gravitational and solar system constraints refer to Eqs. (\ref{eq:ellipse constraint}), (\ref{eq:gravitational constraint}) and (\ref{eq:solar system constraint}), respectively.
(Note that since the solar system constraint is difficult to satisfy, the yellow and orange regions don't appear here.)}
\label{fig:constraint-plot-1}
\end{figure}

The results for the particular case $\beta=1.5$ are shown in Fig. \ref{fig:constraint-plot-1}, where each color corresponds to a specific combination of constraints that are satisfied. 
Note that all constraints are satisfied simultaneously in a few realizations only, represented by the white pixels in the figure. These constitute $0.29\%$ of the total number of simulations with $\beta=1.5$. The percentage obtained for the other values of $\beta$ in the range (\ref{eq:beta range}) is even smaller.

\begin{figure}[h!]
    \centering
    \includegraphics[width=0.49\linewidth]{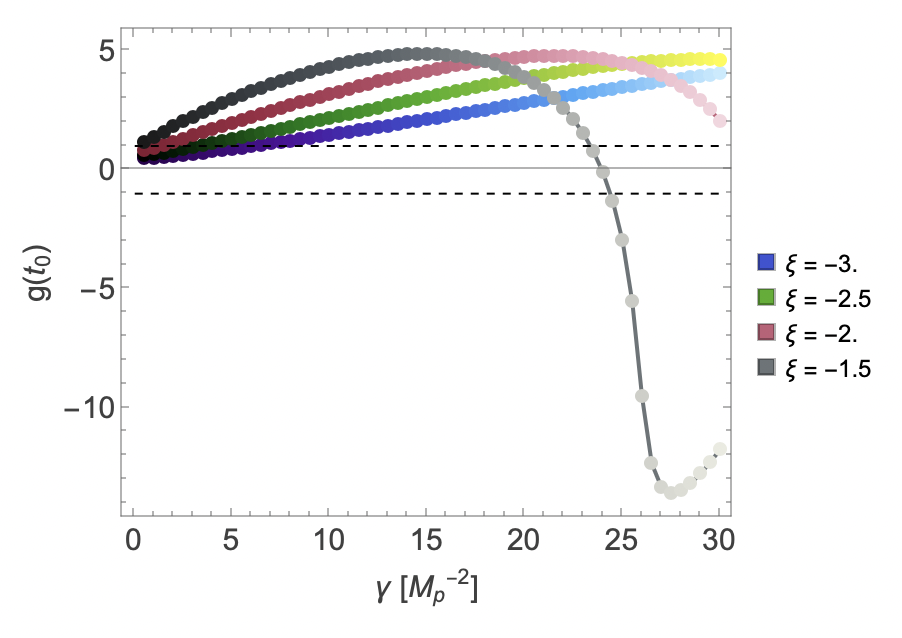}
    \includegraphics[width=0.49\linewidth]{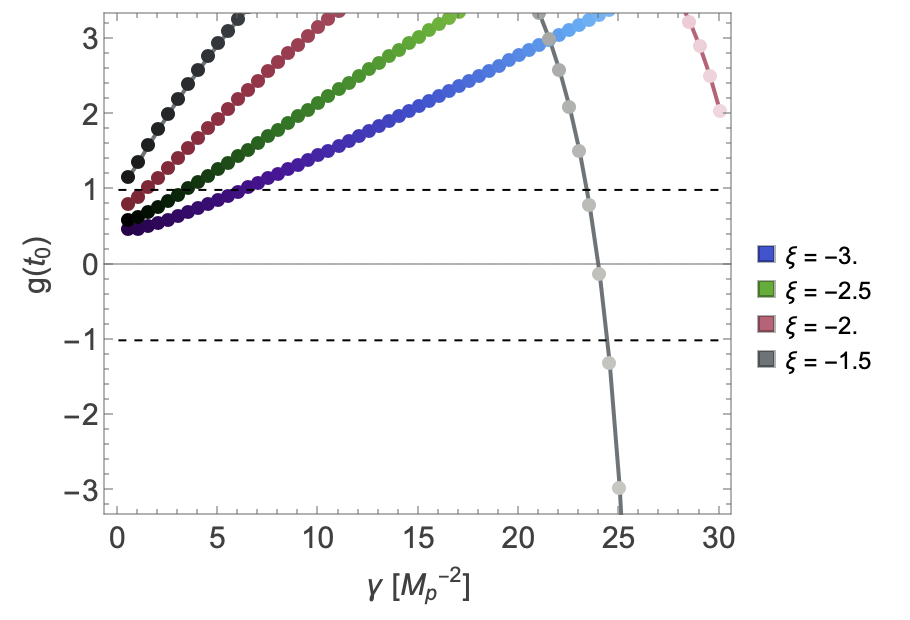}
    \includegraphics[width=0.49\linewidth]{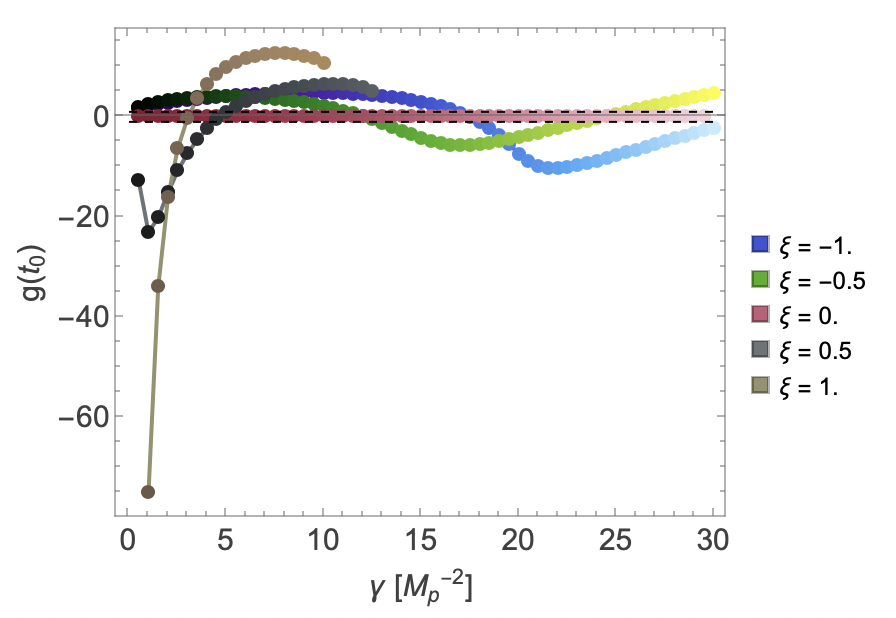}
    \includegraphics[width=0.49\linewidth]{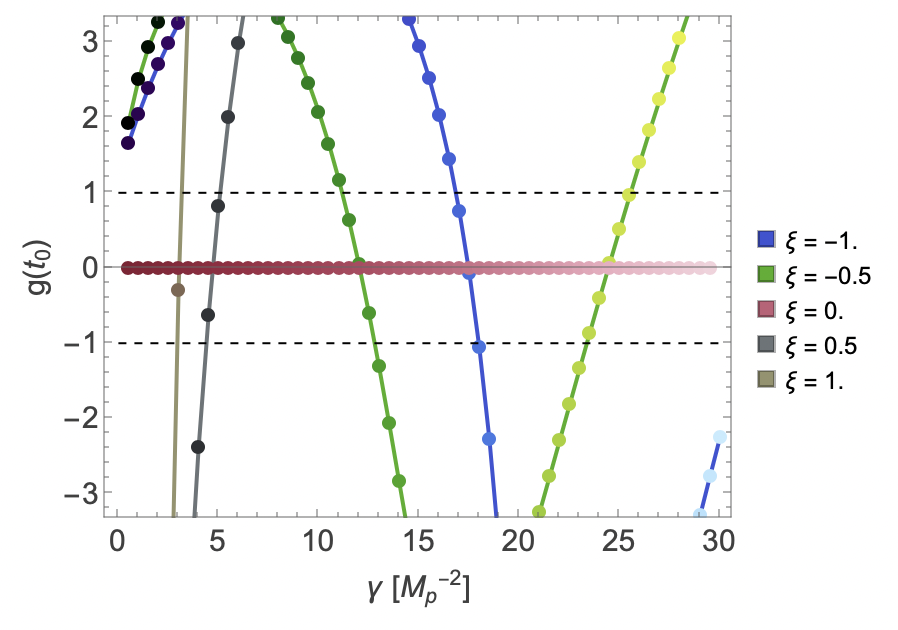}
    \caption{Current value of the function $g(\tilde{t})$ (see Eq. (\ref{eq:g of t})) plotted as a result of varying $\tilde{\gamma}$ from 0.5 to 30 and $\xi$ from -3 to 1, with a fixed $\beta$ of 1.5. The points within each color are varying $\tilde\gamma$ values in steps of 0.5, and the horizontal dashed lines indicate the gravitational constraint bounds (see Eq. (\ref{eq:gravitational constraint})). The panels on the right are zoomed-in versions of the plots on the left. 
    }
    \label{fig:g_constraint-plot}
\end{figure}
\begin{figure}[h!]
    \centering
    \includegraphics[width=0.49\linewidth]{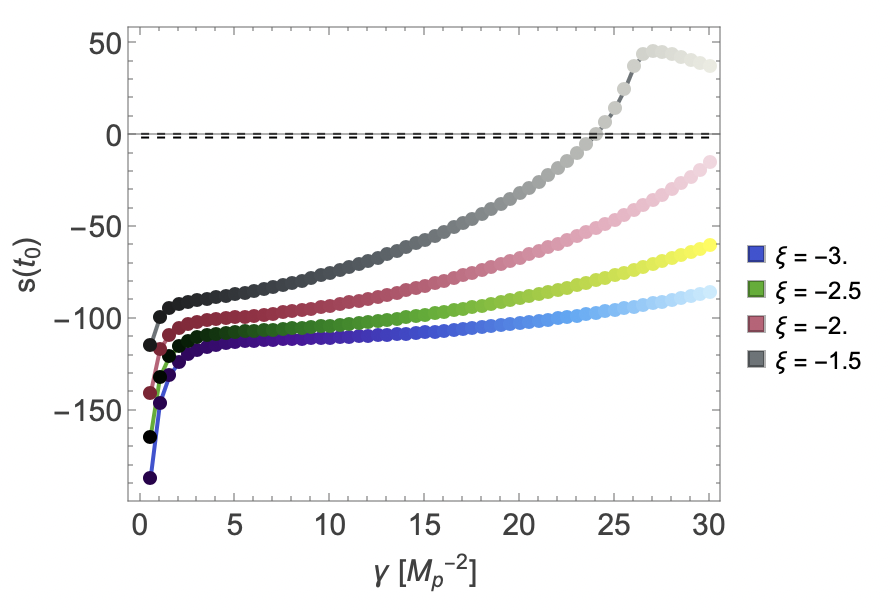}
    \includegraphics[width=0.49\linewidth]{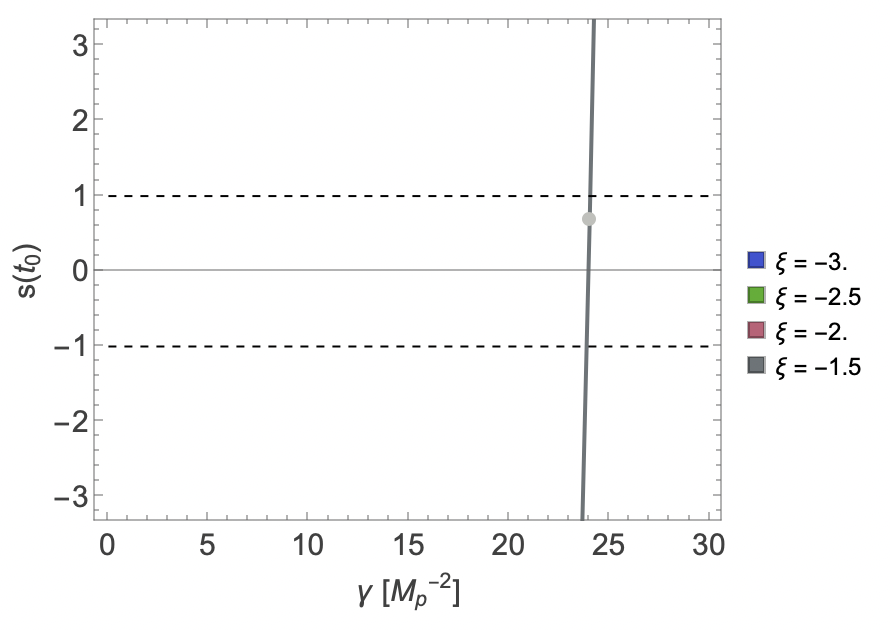}
    \includegraphics[width=0.49\linewidth]{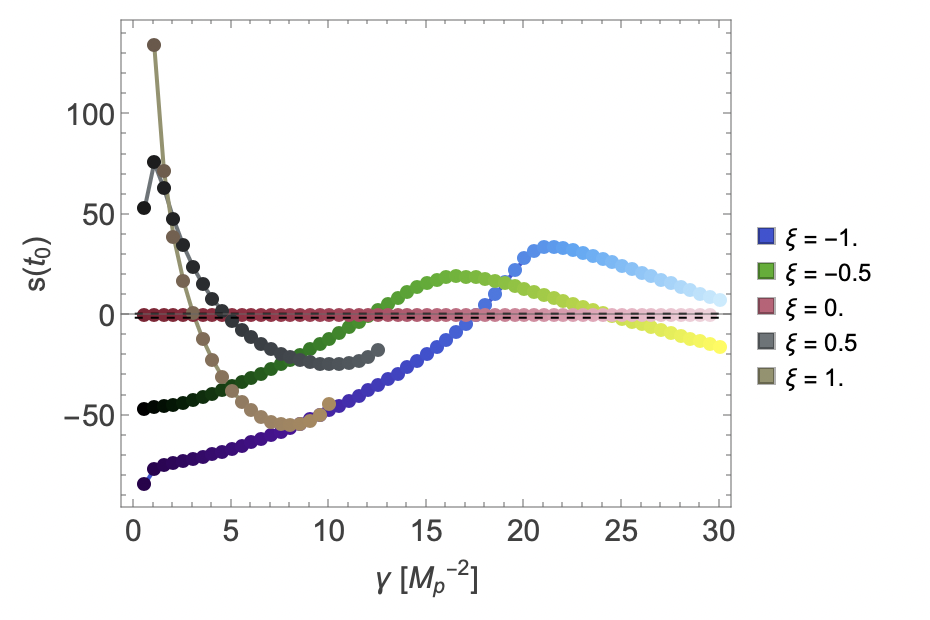}
    \includegraphics[width=0.49\linewidth]{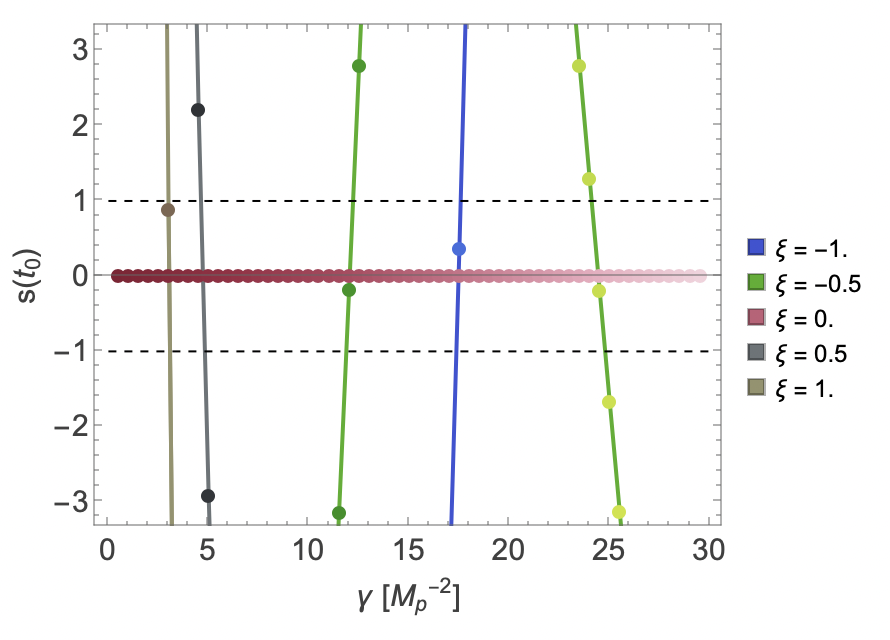}
    \caption{Current value of the function $s(\tilde{t})$ (see Eq. (\ref{eq:s of t})) plotted as a result of varying $\tilde{\gamma}$ from 0.5 to 30 and $\xi$ from -3 to 1, with a fixed $\beta$ of 1.5. The points within each color are varying $\tilde\gamma$ values in steps of 0.5, and the horizontal dashed lines indicate the solar system constraint bounds (see Eq. (\ref{eq:solar system constraint})). The panels on the right are zoomed-in versions of the plots on the left.
    }
    \label{fig:ss_constraint-plot}
\end{figure}
\begin{figure}[h!]
    \centering
    \includegraphics[width=0.49\linewidth]{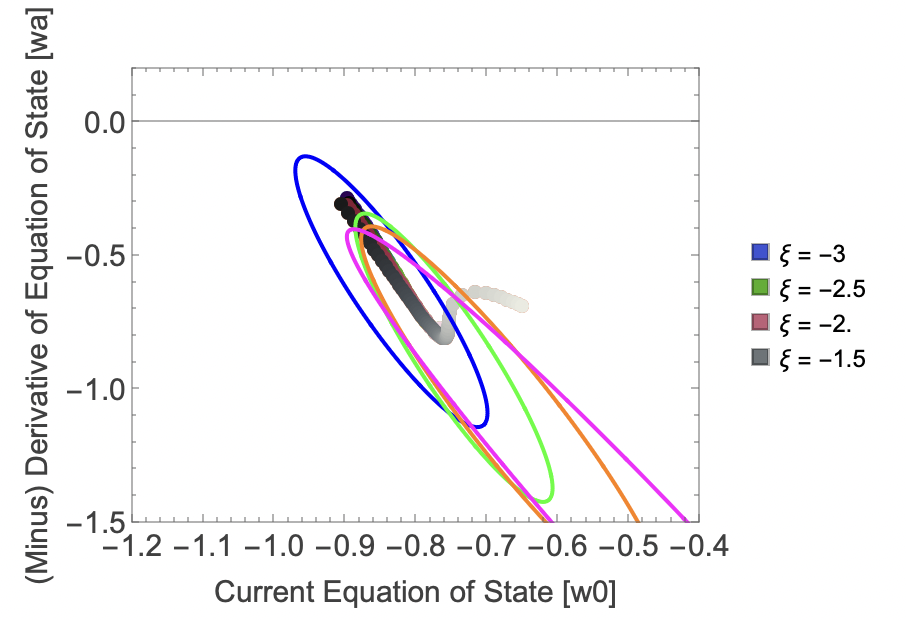}
    \includegraphics[width=0.49\linewidth]{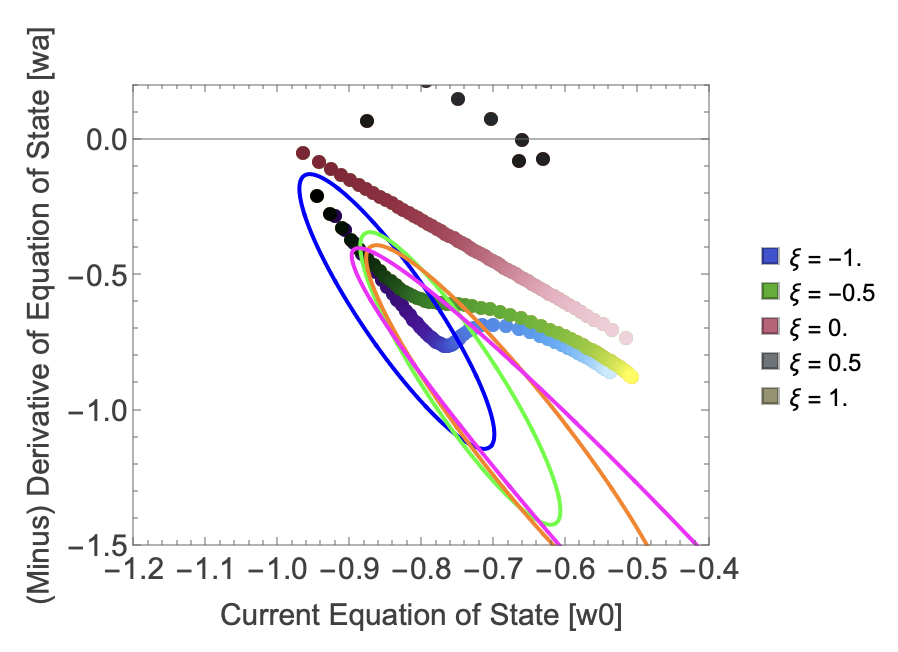}
    \caption{Points on the $w_0,w_a$ plane plotted as a result of varying $\tilde\gamma$ from 0.5 to 30 and $\xi$ from -3 to 1, with a fixed $\beta$ of 1.5. The points within each color are varying $\tilde\gamma$ values in steps of 0.5.}
    \label{fig:w0_wa_sweeps}
\end{figure}

In Figs. \ref{fig:g_constraint-plot} and \ref{fig:ss_constraint-plot}, we show explicitly how the gravitational and solar system constraints are satisfied in this case ($\beta=1.5$) for different values of $\xi$ and $\tilde\gamma$. 
Generically, we find that it is the solar system constraint that is hardest to satisfy (see Table \ref{tab:per_constraints} for the fraction of realizations that satisfy these constraints). 
Note that for $\xi>0$, the curves stop around $\tilde{\gamma}\approx10$; this is because the dark energy density fraction $\bar{\Omega}_{\varphi}$ never reaches 0.69 for higher values of $\tilde{\gamma}$, in agreement with Fig. \ref{fig:constraint-plot-1}.  
We also show in Fig. \ref{fig:w0_wa_sweeps} the corresponding CPL-fitted equation of state parameter (as given by Eq. (\ref{eq:modified w})) in the $w_{0},w_{a}$ plane together with the DESI contours.

\begin{figure}[h!]
    \centering
    \includegraphics[width=0.49\linewidth]{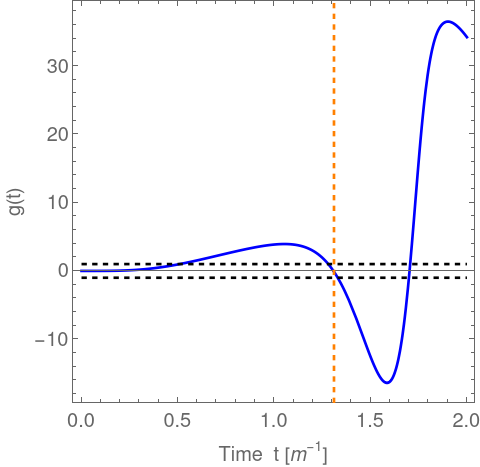}
    \includegraphics[width=0.49\linewidth]{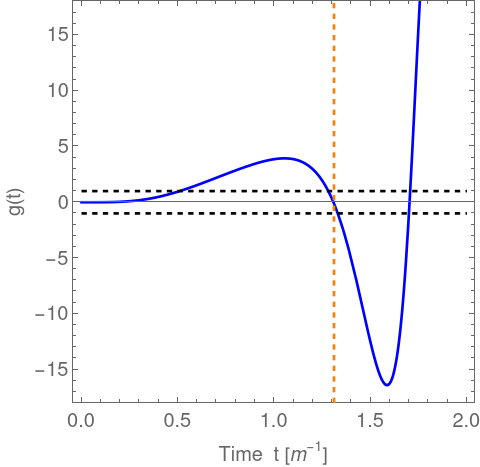}
    \includegraphics[width=0.49\linewidth]{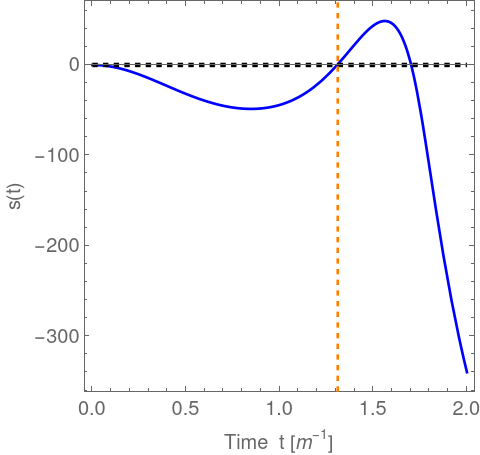}
    \includegraphics[width=0.49\linewidth]{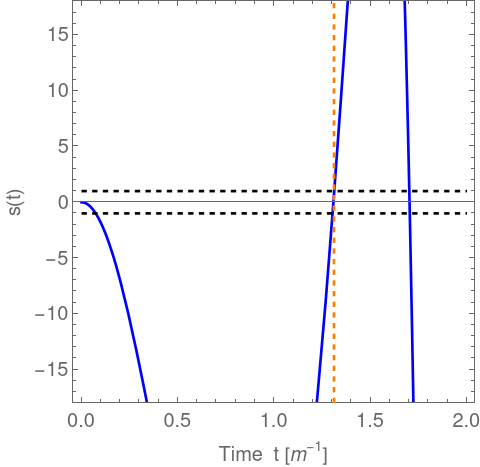}
    \caption{Functions $g(\tilde{t})$ and $s(\tilde{t})$ (see Eqs. (\ref{eq:g of t}) and (\ref{eq:s of t})) in a model with $\beta=1.5$, $\xi=-1.5$ and $\tilde\gamma=24$. The bounds (\ref{eq:gravitational constraint}) and (\ref{eq:solar system constraint}) are indicated by the black dashed lines, and the current time corresponds to the orange dashed line. The plots on the right are zoomed-in versions of the plots on the left. In this model, the constraints are satisfied (compare with Fig. \ref{fig:g and s 2}).}
    \label{fig:g and s 1}
\end{figure}

\begin{figure}[h!]
    \centering
    \includegraphics[width=0.49\linewidth]{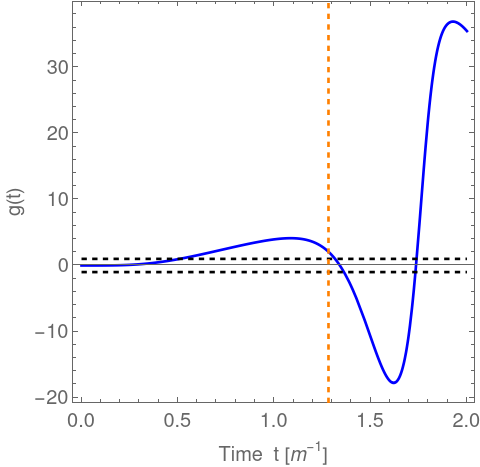}
    \includegraphics[width=0.49\linewidth]{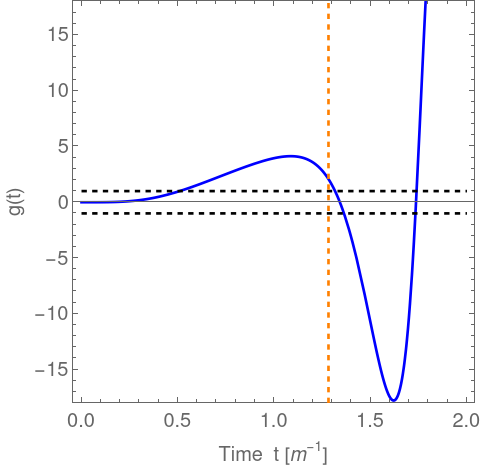}
    \includegraphics[width=0.49\linewidth]{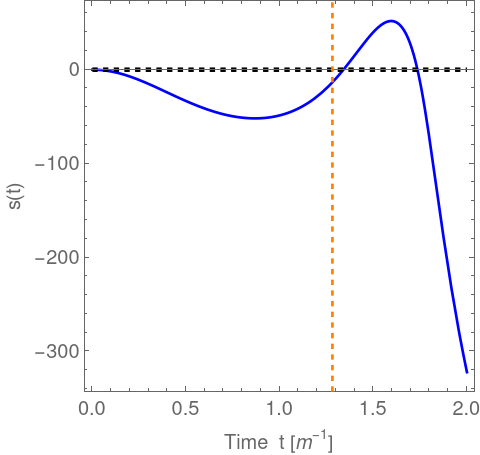}
    \includegraphics[width=0.49\linewidth]{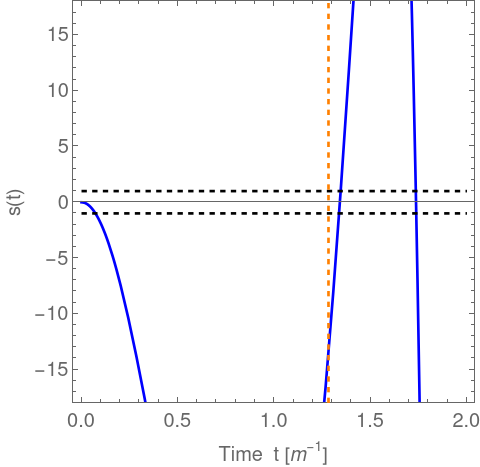}
    \caption{Functions $g(\tilde{t})$ and $s(\tilde{t})$ (see Eqs. (\ref{eq:g of t}) and (\ref{eq:s of t})) in a model with $\beta=1.5$, $\xi=-1.6$ and $\tilde\gamma=24$. The bounds (\ref{eq:gravitational constraint}) and (\ref{eq:solar system constraint}) are indicated by the black dashed lines, and the current time corresponds to the orange dashed line. The plots on the right are zoomed-in versions of the plots on the left. In this model, the constraints are not satisfied (compare with Fig. \ref{fig:g and s 1}).}
    \label{fig:g and s 2}
\end{figure}

According to Eqs. (\ref{eq:g of t}) to (\ref{eq:solar system constraint}), compliance with the gravitational and solar system constraints heavily relies on the smallness of $\varphi$ today. This happens accidentally for the example shown in Fig. \ref{fig:more plots}, where the choice of parameters was $\beta=1.5$, $\xi=-1.5$ and $\tilde\gamma=24$. However, if one chooses slightly different values, $\varphi$ may no longer be sufficiently close to 0 today to satisfy the constraints. We illustrate this point in Figs. \ref{fig:g and s 1} and \ref{fig:g and s 2}, where we show the time evolution of $g(\tilde{t})$ and $s(\tilde{t})$ for two models with the same values of $\beta$ and $\tilde{\gamma}$ (1.5 and 24, respectively), but slightly different $\xi$ (-1.5 for one of the models and -1.6 for the other). While the model with $\xi=-1.5$ is viable, the one with $\xi=-1.6$ is not because $g(\tilde{t}_{0})$ and $s(\tilde{t}_{0})$ shift to new values that are no longer within the observational bounds. 

\begin{table}[H]
\centering
\begin{tabular}{|c|c|c|c|}
\hline
$\xi$ & Solar System & Gravitational & Both \\
\hline
$-3.0$ & 0\% & 19.47\% & 0\% \\
$-2.5$ & 0\% & 9.88\% & 0\% \\
$-2.0$ & 0\% & 3.02\% & 0\% \\
$-1.5$ & 0.61\% & 3.44\% & 0.61\% \\
$-1.0$ & 0.73\% & 3.88\% & 0.73\% \\
$-0.5$ & 3.42\% & 12.9\% & 3.42\% \\
$0$ & 100\% & 100\% & 100\% \\
$0.5$ & 1.62\% & 5.7\% & 1.62\% \\
$1.0$ & 0.81\% & 2.32\% & 0.81\% \\
\hline
\end{tabular}
\caption{Percentage of $\tilde\gamma$ values satisfying each constraint for fixed $\beta$ and varying $\xi$. The $\xi = 0$ case trivially satisfies all constraints. We note that the solar system constraint is rather more restrictive than the gravitational one. This could be altered if the bound on the time variation in Newton's constant is taken to be tighter than $10^{-12}\,\mbox{yr}^{-1}$.}
\label{tab:per_constraints}
\end{table}

\clearpage

\section{Conclusions}
\label{sec:conclusions}
In this work, we have analyzed a specific model of scalar-tensor dark energy with nonminimal coupling in light of the latest DESI data, which suggests that the dark energy density may be evolving with time. 

First, we have revisited the model studied in \cite{Adam:2025kve}, where the scalar field was nonminimally coupled to gravity through a term of the form $\xi\varphi^{2}R$ in the Lagrangian. With a significant fine-tuning of the coupling constant $\xi$ and the initial conditions, that model not only provided a good fit to the DESI data for the dark energy equation of state, but also complied with the gravitational and solar system bounds imposed by several tests of gravity. However, if such a theory is applied directly at very early times, it can lead to unacceptable behavior, with dark energy potentially dominating at earlier times. In this paper, we have clarified the origin of this issue: in general, a scalar field with negative nonminimal coupling to gravity is effectively tachyonic in a matter-dominated background, leading to a rapid growth of the dark energy density toward higher redshifts. To overcome this problem, we extended the framework by introducing a polynomial coupling $\xi f(\varphi)R$, with $f(\varphi)$ given by (\ref{eq:f}), and by selecting initial conditions in which the field is located at the local minimum of the corresponding effective potential. At early times, this is determined by the maximum of $f(\varphi)$. This modification stabilizes the early-time evolution and ensures that the dark energy density remains subdominant throughout the standard cosmological history.

We have explored in a systematic way the parameter space defined by $\xi$, the quartic coupling $\tilde\gamma$ and the slope of the potential $\beta$. The predictions of the model were confronted with the DESI 2025 data on the dark energy equation of state, as well as with constraints from local tests of gravity (in particular, solar system bounds and limits on the time variation of the gravitational coupling). We found that satisfying these constraints and providing a good fit to the DESI data simultaneously is highly non-trivial. Only a tiny fraction of the parameter space explored in this work is viable, forming a narrow strip in the region of negative $\xi$. For the specific example shown in the main text, this fraction was approximately $0.3\%$ in the $\tilde\gamma,\xi$ plane for $\beta=1.5$. 

In the viable regions, the value of the scalar field today is suppressed, which allows the model to evade the constraints. At the same time, the nonminimal coupling plays a crucial role in improving the fit to the DESI data, as it enables phantom behavior of the dark energy equation of state.

Overall, while the extended model resolves the early-time instability and remains compatible with current observational constraints, it does so only in a very restricted region of parameter space, highlighting the degree of tuning still required in this class of scalar-tensor theories.

One way to avoid the solar system tests is to build a model in which the scalar $\varphi$ only couples to the dark matter, rather than the entire matter sector as we have studied here. This would ensure there is no fifth force on regular matter in the solar system. While this idea is interesting, it potentially brings into question the lightness of the scalar's mass; a universally coupled scalar has a somewhat stable mass against renormalization, while a non-universally coupled scalar can have a less stable mass \cite{Hertzberg:2018suv}. In any case, an exploration into a larger class of models seems worthwhile.

\section{Acknowledgments}

M.~P.-H. is supported in part by National Science 
Foundation grants PHY-2310572 and PHY-2419848.
D.~J.-A. is supported in part by National Science 
Foundation grant 
PHY-2419848.
We thank Fabrizio Rompineve for discussion and the VERSE program at Tufts for support. 


\bibliography{DESI2025-bibliography}


\end{document}